\documentclass[a4paper,12pt,english]{article}
\usepackage{amsfonts}
\usepackage{graphicx}
\usepackage{amsmath}
\usepackage{color}

\newcommand{\be}{\begin{equation}}
\newcommand{\ee}{\end{equation}}
\input{tcilatex}

\begin{document}

\begin{titlepage}
\begin{center}

\noindent{{\LARGE{Disk one-point function for a family of non-rational conformal theories}}}

\smallskip
\smallskip

\smallskip
\smallskip
\smallskip
\smallskip
\noindent{\large{Juan Pablo Babaro $^a$ and Gaston Giribet $^{a,b}$}}

\smallskip
\smallskip

\end{center}
\smallskip
\smallskip
\centerline{$^a$ Departamento de F\'{\i}sica, Universidad de Buenos Aires FCEN-UBA}
\centerline{{\it Ciudad Universitaria, Pabell\'on 1, 1428, Buenos Aires, Argentina.}}
\smallskip
\smallskip
\smallskip
\smallskip
\centerline{$^b$ Instituto de F\'{\i}sica de Buenos Aires IFIBA-CONICET}
\centerline{{\it Ciudad Universitaria, Pabell\'on 1, 1428, Buenos Aires, Argentina.}}
\smallskip

\smallskip

\begin{abstract}

We consider an infinite family of non-rational conformal field theories in
the presence of a conformal boundary. These theories, which have
been recently proposed in \cite{ribault1}, are parameterized by two real
numbers $(b,m)$ in such a way that the corresponding central charges $%
c_{(b,m)}$ are given by $c_{(b,m)}=3+6(b+b^{-1}(1-m))^{2}$. For the disk
geometry, we explicitly compute the expectation value of a bulk vertex
operator in the case $m\in \mathbb{Z}$, such that the
result reduces to the Liouville one-point function when $m=0$. We
perform the calculation of the disk one-point function in two different
ways, obtaining results in perfect agreement, and giving the details of
both the path integral and the free field derivations.

\end{abstract}

\end{titlepage}

\newpage

\tableofcontents

\newpage

\section{Introduction}

Non-rational two-dimensional conformal field theories have important
applications in physics. Systems of condensed matter, lower-dimensional
quantum gravity, and string theory are examples of this. In particular, the
problem of considering non-rational models on surfaces with boundaries has
direct applications to the description of D-branes in string theory on
non-compact backgrounds.

The study of conformal field theory (CFT) on two-dimensional manifolds with
boundaries was initiated by J. Cardy in his early work on minimal models 
\cite{Cardy, Cardy2, Cardy3}, and more recently it was extended to
non-rational models by the experts, notably by V. Fateev, A. Zamolodchikov
and Al. Zamolodchikov \cite{FZZ,FZZ2}. The CFT description of branes in
two-dimensional string theory, both in tachyonic and black hole backgrounds,
and in three-dimensional string theory in Anti-de Sitter space (AdS),
attracted the attention of the string theory community in the last ten
years. The literature on string (brane) theory applications of boundary
conformal field theory is certainly vast, and we cannot afford to give a
complete list of references herein. Instead, let us address the reader's
attention to the reviews \cite{Lectures, Lectures2, Lectures3, Yu}.

In this paper, we will consider the family of $c>1$ non-rational
two-dimensional conformal field theories recently proposed in \cite{ribault1}%
. This family of theories is parameterized by two real numbers $(b,m)$ in
such a way that the corresponding central charges $c_{(b,m)}$ are given by $%
c_{(b,m)}=3+6(b+b^{-1}(1-m))^{2}$.

The existence of this family of non-rational models was conjectured in Ref. 
\cite{ribault1} by S. Ribault, who presented convincing evidence by
performing consistency checks. The way of introducing the new CFTs in \cite%
{ribault1} was constructive: These CFTs are defined in terms of their
correlation functions which, on the other hand, admit to be expressed in
terms of correlation functions in Liouville field theory. In addition, a
Lagrangian representation of these new models was also proposed (see (\ref%
{Sbulk}) below), and this was employed in \cite{giribetnakayamanicolas} to
compute two- and three-point functions on the topology of the sphere.

It was emphasized in \cite{ribault1} that, if these non-rational theories
actually exist, then it would be possible to study issues like their
solutions on Riemann surfaces with boundaries. And this is precisely the
task we undertake in this paper: For the disk geometry, and imposing
maximally symmetric boundary conditions, we explicitly compute the
expectation value of a bulk vertex operator in the case $m\in \mathbb{Z}$.
In the case $m=0$ our result reduces to the Liouville disk one-point
function. For the cases $m=1$ and $m=b^{2}$ the observable we compute
corresponds to the one-point function in the SL$(2,\mathbb{C})/$SU$(2)$
Wess-Zumino-Novikov-Witten theory (WZNW), which was recently reconsidered in 
\cite{ribault2} in the context of AdS$_{3}$ string theory. In fact, our
calculation generalizes the calculation of \cite{ribault2} to all values of $%
m\in \mathbb{Z}$, showing that the boundary action for Euclidean AdS$_{2}$
D-branes in Euclidean AdS$_{3}$ space proposed in \cite{ribault2} can be
extended to the whole family of non-rational models of \cite{ribault1}. This
is the case even when the theory for generic $m$ does not exhibit $\widehat{%
sl}(2)_{k}\times \widehat{sl}(2)_{k}$ affine symmetry. Consequently, the
implementation of conformal invariant boundary conditions requires a
condition on the stress-tensor explicitly. The boundary conditions are
discussed in Section 3.

We derive an expression for the one-point function for all values $m\in 
\mathbb{Z}$. The general expression is actually obtained by deriving the
following relation between the disk one-point function in $\mathcal{T}_{b,m}$
and the one-point function in Liouville field theory

\begin{equation}
\bigg\langle\Phi _{j}(\mu |z)\bigg\rangle_{\text{CFT with }c_{(b,m)}}=\delta
^{2}(\mu +\bar{\mu})|\mu |^{m(1+\frac{1-m}{b^{2}})}\,|z-\overline{z}|^{m(1+%
\frac{1}{b^{2}}-\frac{m}{b^{2}})}\bigg\langle V_{\alpha =b(j+1)+\frac{m}{2b}%
}(z)\bigg\rangle_{\text{Liouville}}  \label{torbe}
\end{equation}%
where the observable on the left hand side corresponds to the expectation
value of a primary bulk operator $\Phi _{j}(\mu |z)$ of conformal dimension $%
h_{j}=-b^{2}j(j+1)+(j+1)(1-m)$ in the CFT of central charge $c_{(b,m)}\ $%
formulated on the disk geometry (see \cite{ribault1} for details). On the
other hand, the observable on the right hand side corresponds to the
one-point function of an exponential vertex operator of momentum $\alpha
=b(j+1)+m/(2b)$ in Liouville theory on the disk.

To complete the statement in (\ref{torbe}), we have also to give the
definition of the boundary parameter for the one-point function in $\mathcal{%
T}_{b,m}$. In Liouville theory, the one-point function $\left\langle
V_{\alpha }(z)\right\rangle _{\text{Liouville}}$ is parameterized by a real
variable $s$ \cite{teschner}; in turn, it is necessary to complement (\ref%
{torbe})\ by introducing a boundary parameter $r$ in $\mathcal{T}_{b,m}$ and
specifying how it relates to the Liouville parameter $s$. Such relation
reads 
\begin{equation}
r=2\pi bs+i\frac{m\pi }{2}\text{sgn}(\text{Im}\mu ).  \label{asdefined}
\end{equation}%
The need of introducing this boundary parameter $r$ in $\mathcal{T}_{b,m}$
comes from the following fact:\ In Liouville theory, parameter $s$ is
determined by the boundary cosmological constant $\mu _{B}$ appearing in the
action only up to shifts $s\rightarrow s+ib^{-1}n$ with $n\in \mathbb{Z}$
(see (\ref{L1})-(\ref{L3}) below), and some observables of the theory, such
as the Liouville one-point function, depend on $s$ in a way that is not
invariant under such shifts. Therefore, to completely specify the relation
between the left hand side and the right hand side of (\ref{torbe}) it is
necessary to prescribe (\ref{asdefined}) and then give a definition for the
boundary parameter in $\mathcal{T}_{b,m}$ in terms of $s$. In the path
integral computation we will perform in Section 4, the parameter $r$ will be
ultimately related to the coupling constant $\xi $ appearing in the action
(see (\ref{siete}) below).

In this paper we will give two different proofs of (\ref{torbe}) for the case%
\footnote{%
It seems natural to conjecture that a relation like (\ref{torbe}) with (\ref%
{asdefined}) also holds for arbitrary (not necessarily integer) values of $m$%
.} $m\in \mathbb{Z}$. That is, we perform the calculation of the bulk
one-point function in two different ways:\ First, in Section 4 we give a
path integral derivation. This consists of reducing the calculation of the
expectation value of one bulk operator on the disk geometry to the analogous
quantity in Liouville field theory, which, conveniently, is already known 
\cite{FZZ}. We closely follow the path integral techniques developed in \cite%
{schomerus}, also reproduced in \cite{ribault2, hikidaschomerus2}. With the
intention to make the comparison with these reference easy, we aim to use a
notation similar to the one used therein, up to minor differences. In
Section 5, following the steps of \cite{ribault2}, we perform a free-field
calculations of the same one-point function, finding perfect agreement with
the path integral computation. In Section 6, we analyze some functional
properties of the formula we obtain, of which the one-point function in
Liouville theory and in the WZNW theory are particular cases.

\section{A family of non-rational conformal theories}

The action of the non-rational conformal field theories discussed in \cite%
{ribault1} is given by

\begin{equation}
S_{\text{bulk}}\ =\ \frac{1}{2\pi }\int_{\Gamma }d^{2}z\,g^{1/2}\left(
\partial \phi \bar{\partial}\phi +\beta \bar{\partial}\gamma +\bar{\beta}%
\partial \bar{\gamma}+\frac{Q_{m}}{4}R\phi +b^{2}(-\beta \bar{\beta}%
)^{m}\,e^{2b\phi }\right)  \label{Sbulk}
\end{equation}%
where $Q_{m}=b+b^{-1}(1-m)$, and where we are using the standard notation $%
\partial =\frac{\partial }{\partial z}=\frac{1}{2}\partial _{x}-\frac{i}{2}%
\partial _{y}$ and $\overline{\partial }=\frac{\partial }{\partial \overline{%
z}}=\frac{1}{2}\partial _{x}+\frac{i}{2}\partial _{y}$. We are interested in
the case of the Riemann surface $\Gamma $ being the disk geometry. Conformal
invariance will allow us to represent the region integration $\Gamma $ with
the upper half plane. The notation $S_{\text{bulk}}$ in (\ref{Sbulk}) is
chosen to emphasize that we are not including boundary terms yet (see (\ref%
{conboundaries}) below).

Let us call $\mathcal{T}_{m,b}$ the theory defined by the action (\ref{Sbulk}%
). Then, we notice that the case $\mathcal{T}_{0,b}$ corresponds to
Liouville field theory coupled to a free $\beta $-$\gamma $ ghost system. On
the other hand, the case $\mathcal{T}_{1,1/\sqrt{k-2}}$ corresponds to the $%
H_{3}^{+}$=SL$(2,\mathbb{C})/$SU$(2)$ WZNW theory with level $k=b^{-2}+2$
written in the Wakimoto free-field representation \cite{Wakimoto}. Besides, $%
\mathcal{T}_{k-2,\sqrt{k-2}}$ also yields the $H_{3}^{+}$ WZNW theory, with
level $k=b^{+2}+2$; see \cite{giribetnakayamanicolas}. In fact, the action
above can be regarded as a generalization of these well-known non-rational
conformal field theories.

The free stress-tensor associated to (\ref{Sbulk}) is given by 
\begin{equation}
T(z)\ =\ -\beta (z)\partial \gamma (z)-(\partial \phi
(z))^{2}+Q_{m}\,\partial ^{2}\phi (z)  \label{T}
\end{equation}%
and by its anti-holomorphic counterpart $\overline{T}(\overline{z})$. This
yields the central charge of the theory 
\begin{equation}
c_{(b,m)}=1+6Q_{m}^{2}.  \label{ccc}
\end{equation}%
The conformal dimension of the field $e^{2\alpha \phi }$ with respect to (%
\ref{T}) is $(h_{\alpha },\bar{h}_{\alpha })=(\alpha (Q_{m}-\alpha ),\alpha
(Q_{m}-\alpha ))$, while the conformal dimensions of $\beta $ and $\gamma $
are $(1,0)$ and $(0,0)$, respectively. Therefore, the last term in the (\ref%
{Sbulk}) is marginal with respect to the stress-tensor (\ref{T}) if $h_{b}\
=\ \bar{h}_{b}\ =\ b\,(Q_{m}-b)=1-m$, yielding the relation

\begin{equation}
Q_{m}\ =\ b+\frac{1-m}{b}.
\end{equation}

As mentioned in the Introduction, the theory with $m=1$ (and the theory with 
$m=b^{2}$) corresponds to the SL$(2,\mathbb{C})/$SU$(2)$ WZNW model, which
exhibits $\widehat{sl}(2)_{k}\times \widehat{sl}(2)_{k}$ affine Kac-Moody
symmetry. This symmetry is generated by the Kac-Moody current algebra 
\begin{eqnarray}
J^{-}(z) &=&\beta (z)  \label{J1} \\
J^{3}(z) &=&\beta (z)\gamma (z)+b^{-1}\partial \phi (z) \\
J^{+}(z) &=&\beta (z)\gamma ^{2}(z)+2b^{-1}\gamma (z)\partial \phi
(z)-\left( b^{-2}+2\right) \partial \gamma (z)  \label{J3}
\end{eqnarray}%
together with the anti-holomorphic counterparts $\overline{J}^{3,\pm }(%
\overline{z})$, where $b^{-2}=k-2$. In contrast, the theory (\ref{S}) for
generic $m$ exhibits only the algebra generated only by the two currents $%
J^{3}(z)$ and $J^{-}(z)$, 
\begin{equation}
J^{-}(z)=\beta (z),\qquad \qquad J^{3}(z)=\beta (z)\gamma (z)+\frac{m}{b}%
\partial \phi (z)  \label{14}
\end{equation}%
and by the pair of anti-holomorphic analogues. Currents (\ref{14}) obey the
following operator product expansion (OPE) 
\begin{equation}
J^{-}(z)J^{3}(w)\simeq \frac{J^{-}(w)}{(z-w)}+...\qquad
J^{3}(z)J^{3}(w)\simeq -\frac{(1+m^{2}b^{-2}/2)}{(z-w)^{2}}+...\qquad
J^{-}(z)J^{-}(w)\simeq ...
\end{equation}%
where the ellipses stand for regular terms that are omitted; and this
realizes the Lie brackets for the modes $J_{n}^{3,-}=\frac{1}{2\pi i}\int
dzJ^{3,-}(z)z^{-n-1}$. The spectrum of the theory would be constructed in
terms of primary states with respect to these currents. Vertex operators
creating such states are of the form 
\begin{equation}
\Phi ^{j}(\mu |z)\ =\ |\rho (z)|^{2h_{j}}\,|\mu |^{2m(j+1)}\,e^{\mu \gamma
(z)-\bar{\mu}\bar{\gamma}(\bar{z})}\,e^{2b(j+1)\phi (z,\bar{z})}
\label{CuatrO}
\end{equation}%
whose holomorphic and anti-holomorphic conformal dimensions are given by

\begin{equation}
h_{j}\ =\ \bar{h}_{j}\ =(-b^{2}j+1-m)(j+1),
\end{equation}
and where $\mu $ is a complex variable. The spectrum of normalizable states
of the theory, which is ultimately expressed by the values that $j$ takes,
is to be determined. The dependence on $\rho (z)$ in (\ref{CuatrO}) was
introduced because throughout this paper we will work in the conformal gauge 
$ds^{2}=|\rho (z)|^{2}dzd\bar{z}$, and such dependence is the one required
for $\Phi ^{j}(\mu |z)$ to transform as a primary ($h_{j},\overline{h}_{j}$%
)-dimension operator under conformal transformations.

\section{Boundary action and boundary conditions}

The boundary term we add to the action (\ref{Sbulk}) to define the theory in
a surface with boundaries\ is the following

\begin{equation}
S_{\text{boundary}}\ =\ \frac{1}{2\pi }\int_{\partial \Gamma
}dx\,g^{1/4}\left( Q_{m}K\phi \,+\,\frac{i}{2}\beta \left( \gamma +\bar{%
\gamma}-\xi \,\beta ^{m-1}\,e^{b\phi }\right) \right) ,
\label{conboundaries}
\end{equation}%
where $\xi $ is an arbitrary constant. This is the natural generalization of
the boundary action proposed in \cite{ribault2}. Since the field $\beta $ is
not supposed to be positive, we assume here that $m\in \mathbb{Z}$. This is
a limitation of both our path integral calculation and our free field
calculation, and this is why the expression we get is valid for $m\in 
\mathbb{Z}$.

We will consider the theory on the disk. In turn, conformal invariance
permits to chose $\Gamma $ as being the upper half plane Im$(z)=y\geq 0$,
with $z=x+iy$, and, consequently, the boundary is given by the real line Re$%
(z)=x$. On the disk geometry, we will consider the case of maximally
symmetric boundary conditions (see (\ref{tres})-(\ref{Dos}) below). To
implement this, first we integrate by parts the $\gamma $-$\beta $ terms of (%
\ref{conboundaries}). Then, integrating the resulting expression over the
semi-infinite line Re$(y)>0$, we see that two pieces $-\frac{i}{2}\beta
\gamma |_{y=0}$ and $+\frac{i}{2}\bar{\beta}\bar{\gamma}|_{y=0}$ come to
cancel the contribution $\beta (\gamma +\bar{\gamma})$ in (\ref%
{conboundaries}). The total action then takes the form 
\begin{equation}
\begin{split}
S\ & =\ \frac{1}{2\pi }\int_{\Gamma }d^{2}z\,g^{1/2}\left( \partial \phi 
\bar{\partial}\phi -\gamma \bar{\partial}\beta -\bar{\gamma}\partial \bar{%
\beta}+\frac{Q_{m}}{4}R\phi +b^{2}(-\beta \bar{\beta})^{m}\,e^{2b\phi
}\right) + \\
& +\frac{1}{2\pi }\int_{\partial \Gamma }dx\,g^{1/4}\left( Q_{m}K\phi -\frac{%
i\xi }{2}\beta ^{m}\,e^{b\phi }\right)
\end{split}
\label{S}
\end{equation}

It is convenient to introduce bulk and boundary coupling constants $(\lambda
,\lambda _{B})$ to control the strength of the interacting terms. This is
achieved by shifting the zero-mode of $\phi $ as follows $\phi \rightarrow
\phi +\frac{1}{2b}\ln \left( \lambda /b^{2}\right) $ and redefining $\xi
=2ib\lambda _{B}/\sqrt{\lambda }$. This makes the interaction term in the
action to take the form $\lambda \int_{\Gamma }(-\beta \bar{\beta}%
)^{m}e^{2b\phi }+\lambda _{B}\int_{\partial \Gamma }\beta ^{m}e^{b\phi }$,
which is useful to perform the free field calculation (see Section 5) as in
this way one has access to a perturbative treatment of the screening
effects. The coupling $\xi $ takes imaginary values for the boundary action
to be real.

Varying the action (\ref{S}) in the boundary and imposing the condition $%
\delta (\beta +\bar{\beta})_{|z=\bar{z}}=0$, one finds

\begin{equation}
\delta S|_{\partial \Gamma }\ =\ \frac{i}{4\pi }\int dx\bigg[\left(
(-\partial +\bar{\partial})\phi -\xi \,b\,\beta ^{m}\,e^{b\phi }\right)
\delta \phi +\left( \gamma +\bar{\gamma}-\xi \,m\,\beta ^{m-1}\,e^{b\phi
}\right) \delta \beta \bigg]
\end{equation}%
From this, one reads the gluing conditions in $z=\bar{z}=x$, which have to be

\begin{eqnarray}
\beta (x)+\bar{\beta}(x)\text{ }_{|z=\bar{z}} &=&0  \label{tres} \\
\gamma (x)+\bar{\gamma}(x)\text{ }_{|z=\bar{z}} &=&\xi \,m\,\beta
^{m-1}(x)\,e^{b\phi (x)} \\
(-\partial +\bar{\partial})\,\phi (x)_{\text{ }|z=\bar{z}} &=&\xi \,b\,\beta
^{m}\,(x)e^{b\phi (x)}  \label{Dos}
\end{eqnarray}

And these conditions correspond to 
\begin{eqnarray}
J^{-}(x)+\bar{J^{-}}(x)\ _{|z=\bar{z}} &=&\ 0,  \label{20} \\
J^{3}(x)-\bar{J^{3}}(x)\ _{|z=\bar{z}} &=&\ 0.  \label{21}
\end{eqnarray}

While (\ref{20}) is evidently satisfied, condition (\ref{21}) can be checked
taking into account the conditions (\ref{tres})-(\ref{Dos}).

The other condition to be considered in the boundary is%
\begin{equation}
T\,(x)-\,\bar{T}(x)\ _{|z=\bar{z}}=\ 0
\end{equation}%
which guarantees boundary conformal symmetry. This classical analysis
suggests that with these boundary conditions one obtains a theory that
preserves the conformal symmetry generated by $J^{-}(z)$, $J^{3}(z)$, and $%
T(z)$. The free field computation ultimately helps to prove this explicitly.

Now we have discussed the boundary conditions and proposed the form of the
boundary action, we are ready to undertake the calculation of the one-point
function.

\section{Path integral computation}

The one-point function we are interested in is the vacuum expectation value $%
\left\langle \Phi ^{j}(\mu |z)\right\rangle _{\mathcal{T}_{m,b}}$ of one
bulk vertex operator (\ref{CuatrO}) in the disk geometry. This is given by 
\begin{equation}
\Omega _{j}^{(m,b)}(z):=\left\langle \Phi ^{j}(\mu |z)\right\rangle _{%
\mathcal{T}_{m,b}}=\ \int \mathcal{D}\phi \mathcal{D}^{2}\beta \mathcal{D}%
^{2}\gamma \mathcal{\ }\,e^{-S}\,|\rho (z)|^{2h_{j}}\,|\mu
|^{2m(j+1)}\,e^{\mu \gamma (z)-\bar{\mu}\bar{\gamma}(\bar{z}%
)}\,e^{2b(j+1)\phi (z,\bar{z})},  \label{TheSD}
\end{equation}%
with $\mathcal{D}^{2}\beta =\mathcal{D}\beta \mathcal{D}\bar{\beta}$, and $%
\mathcal{D}^{2}\gamma =\mathcal{D}\gamma \mathcal{D}\bar{\gamma}$, and
imposing $\beta +\bar{\beta}\ =0$ on the line $z=\bar{z}$ as boundary
condition. The action $S$ in (\ref{TheSD}) is given by (\ref{S}).

Integrating over the fields $\gamma $ and $\bar{\gamma}$ we get

\begin{equation}
\int \mathcal{D}\gamma \,e^{\frac{1}{2\pi }\int d^{2}w\gamma \bar{\partial}%
\beta }\,e^{\mu \gamma (z)}\ =\ \delta \left( \frac{1}{2\pi }\,\bar{\partial}%
\beta (w)+\mu \,\delta ^{(2)}(w-z)\right)
\end{equation}%
and, respectively, 
\begin{equation}
\int \mathcal{D}\bar{\gamma}\,e^{\frac{1}{2\pi }\int d^{2}\bar{w}\bar{\gamma}%
\partial \bar{\beta}}\,e^{\bar{\mu}\bar{\gamma}(\bar{z})}\ =\ \delta \left( 
\frac{1}{2\pi }\,\partial \bar{\beta}(\bar{w})-\bar{\mu}\,\delta ^{(2)}(\bar{%
w}-\bar{z})\right) .
\end{equation}%
Using that $\bar{\partial}\left( \frac{1}{z}\right) =\partial \left( \frac{1%
}{\bar{z}}\right) =2\pi \delta ^{(2)}(z)$ we can write 
\begin{equation}
\frac{1}{2\pi}\,\bar{\partial}\beta(w)+\frac{\mu}{2\pi}\,\bar{\partial}\left(%
\frac{1}{w-z}\right)\ =\ 0
\end{equation}

\begin{equation}
\frac{1}{2\pi}\,\partial\bar{\beta}(\bar{w})-\frac{\bar{\mu}}{2\pi}%
\,\partial\left(\frac{1}{\bar{w}-\bar{z}}\right)\ =\ 0
\end{equation}

Then, in the boundary we impose 
\begin{equation}
\beta +\bar{\beta}\ _{|z=\bar{z}}=0.
\end{equation}

Now, we integrate over $\beta $ and $\bar{\beta}$. Considering the
conditions above, we get a non-vanishing solution only if $\mu +\bar{\mu}=0$%
. The solution for $\beta $ and $\bar{\beta}$ are thus given by 
\begin{eqnarray}
\beta _{0}(w)\ &=&\ -\frac{\mu }{w-z}-\frac{\bar{\mu}}{w-\bar{z}}\ =\ \frac{%
-\mu \,(z-\bar{z})}{(w-z)(w-\bar{z})},  \label{cuatro} \\
\bar{\beta}_{0}(\bar{w})\ &=&\ \frac{\mu }{\bar{w}-z}+\frac{\bar{\mu}}{\bar{w%
}-\bar{z}}\ =\ \frac{-\bar{\mu}\,(z-\bar{z})}{(\bar{w}-z)(\bar{w}-\bar{z})}.
\label{cinco}
\end{eqnarray}

Now, we can follow the analysis of \cite{schomerus} closely. We consider the
exact differential $\rho (w)\beta _{0}(w)$, whose expression we can give in
terms of its poles, up to a global factor $u$. Then, evaluating $\beta $ the
function to be computed can be seen to take the form 
\begin{equation}
\begin{split}
& \Omega _{j}^{(m,b)}(z)=\int \mathcal{D}\phi \exp -\left( \frac{1}{2\pi }%
\int d^{2}w\left( \partial \phi \bar{\partial}\phi
\,+\,b^{2}|u|^{2m}|w-z|^{-2m}|w-\bar{z}|^{-2m}\,|\rho (w)|^{-2m}\,e^{2b\phi
}\right) \right) \times \\
& \times \exp \left( -\frac{i\xi }{4\pi }\int d\tau \,u^{m}(\tau
-z)^{-m}\,(\tau -\bar{z})^{-m}\,\rho (w)^{-m}\,e^{b\phi }\right) |\rho
(z)|^{2h_{j}}\,|\mu |^{2m(j+1)}\,e^{2b(j+1)\phi (z,\overline{z})}\delta
^{(2)}(\mu +\bar{\mu})
\end{split}%
\end{equation}

Now, consider the following change of variables (i.e. field redefinition) 
\begin{equation}
\phi (w)\rightarrow \phi (w)-\frac{m}{b}\ln |u|,
\end{equation}%
which consequently gives $e^{b\phi (w)}\rightarrow e^{b\phi (w)}\,|u|^{-m}$.
This induces a change in the linear \textquotedblright dilaton" term $-\frac{%
1}{8\pi }\int_{\Gamma }d^{2}w\,g^{1/2}\,Q_{m}R\phi \,-\,\frac{1}{2\pi }%
\int_{\partial \Gamma }d\tau \,g^{1/4}KQ_{m}\phi $, which now takes the form%
\begin{eqnarray}
&&-\frac{1}{8\pi }\int d^{2}w\,g^{1/2}\,Q_{m}R\phi \,-\,\frac{1}{2\pi }%
\int_{\partial \Gamma }d\tau \,g^{1/4}KQ_{m}\phi \,+  \notag \\
&&+\,\ln |u|m\left( 1+\frac{1-m}{b^{2}}\right) \left[ \frac{1}{8\pi }%
\int_{\Gamma }d^{2}w\,g^{1/2}R+\frac{1}{2\pi }\int_{\partial \Gamma }d\tau
\,g^{1/4}K\right]
\end{eqnarray}

Using the Gauss-Bonnet theorem, which states that the Euler characteristic
of the disk is 
\begin{equation}
\chi (\Gamma )\ =\frac{1}{8\pi }\int_{\Gamma }d^{2}w\,g^{1/2}R\,+\,\frac{1}{%
2\pi }\int_{\partial \Gamma }d\tau \,g^{1/4}K\ =1,
\end{equation}
we find 
\begin{equation}
\begin{split}
\Omega _{j}^{(m,b)}(z)=\int \mathcal{D}\phi \exp -\left( \frac{1}{2\pi }\int
d^{2}w\left( \partial \phi \bar{\partial}\phi +b^{2}|w-z|^{-2m}\,|w-\bar{z}%
|^{-2m}|\rho (z)|^{-2m}e^{2b\phi }\right) \right) \times \\
\times \exp \left( -\frac{i\xi }{4\pi }\left( \frac{u}{|u|}\right) ^{m}\int
d\tau (\tau -z)^{-m}\,(\tau -\bar{z})^{-m}\,\rho ^{-m}(\tau )e^{b\phi
}\right) \times \\
\times \delta ^{(2)}(\mu +\bar{\mu})|\rho (z)|^{2h_{j}}\,|\mu
|^{2m(j+1)}\,|u|^{-m(2j+1+b^{-2}\left( m-1\right) )}e^{2b(j+1)\phi (z,\bar{z}%
)}
\end{split}%
\end{equation}

Now, let us perform a second change of variables, 
\begin{equation}
\phi (w,\bar{w})=\varphi (w,\bar{w})+\frac{m}{2b}\left( \ln |w-z|^{2}+\ln |w-%
\bar{z}|^{2}+\ln |\rho (w)|^{2}\right) ,  \label{ASD}
\end{equation}%
which amounts to say 
\begin{equation}
e^{b\phi }=e^{b\varphi }\,|w-z|^{m}\,|w-\bar{z}|^{m}\,|\rho (w)|^{m}
\end{equation}%
and 
\begin{equation}
\partial \bar{\partial}\phi (w,\bar{w})=\partial \bar{\partial}\varphi (w,%
\bar{w})+\frac{m\pi }{b}\delta ^{(2)}\left( |w-z|\right) +\frac{m\pi }{b}%
\delta ^{(2)}(|w-\bar{z}|)+\frac{m}{2b}\partial \bar{\partial}\ln |\rho
(w)|^{2}.
\end{equation}%
In the bulk action this change produces a transformation in the kinetic term 
$-\frac{1}{2\pi }\int_{\Gamma }d^{2}w\partial \phi \bar{\partial}\phi $,
which becomes 
\begin{eqnarray}
&&\frac{1}{2\pi }\int_{\Gamma }d^{2}w\,\varphi \partial \bar{\partial}%
\varphi +\frac{m}{2b}\int_{\Gamma }d^{2}w\,\varphi \,\delta ^{(2)}(|w-z|)+%
\frac{m}{2b}\int_{\Gamma }d^{2}w\,\varphi \,\delta ^{(2)}(|w-\bar{z}|)+ 
\notag \\
&&+\frac{m}{4\pi b}\int_{\Gamma }d^{2}w\,\varphi \,\partial \bar{\partial}%
\ln |\rho (w)|^{2}+\frac{m^{2}}{4\pi b^{2}}\int_{\Gamma }d^{2}w\left( \ln
|w-z|^{2}+\ln |w-\bar{z}|^{2}+\ln |\rho (w)|^{2}\right) \times  \notag \\
&&\times \left( \frac{b}{m}\partial \bar{\partial}\varphi +\pi \delta
^{(2)}(|w-z)|+\pi \delta ^{(2)}(|w-\bar{z}|)+\frac{1}{2}\,\partial \bar{%
\partial}\ln |\rho (w)|^{2}\right) .  \label{50}
\end{eqnarray}

Using the regularization \cite{schomerus} 
\begin{equation}
\lim_{w\rightarrow z}\ln |w-z|^{2}\equiv \ln |\rho (z)|^{2},  \label{L42}
\end{equation}%
being 
\begin{equation}
ds^{2}\ =\ |\rho (z)|^{2}\,dzd\bar{z},\qquad \sqrt{g}R\ =-4\,\partial \bar{%
\partial}\ln |\rho (z)|^{2},
\end{equation}%
one finds that the right hand side of (\ref{50}) takes the form%
\begin{equation}
\ -\,\frac{1}{2\pi }\int_{\Gamma }d^{2}w\,\partial \varphi \bar{\partial}%
\varphi \,-\,\frac{1}{2\pi }\int_{\Gamma }d^{2}w\,\sqrt{g}\frac{m}{4b}%
R\varphi \,+\frac{m}{b}\varphi (z)+\frac{m^{2}}{2b^{2}}\ln |z-\overline{z}|+%
\frac{m^{2}}{4b^{2}}\ln |\rho (z)|^{2}.
\end{equation}

The boundary action also suffers a change. Recalling how to write the
extrinsic curvature $K$ in terms of $\partial _{x}^{2}\ln |\rho (\tau )|^{2}$%
, we find that, under the change (\ref{ASD}), the boundary action changes as
follows 
\begin{equation}
\delta S_{\text{boundary}}=\ \frac{m}{2\pi b}\int_{\partial \Gamma
}\,g^{1/4}K\phi .
\end{equation}%
In turn, the changes above induce a modification in the value of the
background change $Q_{m}$, shifting it as follows 
\begin{equation}
Q_{m}\ \rightarrow \ Q_{m}+\frac{m}{b}\ =\ b+\frac{1-m}{b}+\frac{m}{b}\ =\ b+%
\frac{1}{b}\ =\ Q_{m=0}\ 
\end{equation}%
which gives the Liouville background charge $Q_{\text{Liouville}}=b+1/b$ as
the result.

With all this, the one-point function we are trying to compute takes the
form 
\begin{equation}
\begin{split}
\Omega _{j}^{(m,b)}(z)& =\ \delta ^{2}(\mu +\bar{\mu})|u|^{m(1+\frac{1-m}{%
b^{2}})}\,|z-\bar{z}|^{\frac{m^{2}}{2b^{2}}}\int \mathcal{D}\varphi \exp %
\left[ -\left( \frac{1}{2\pi }\int_{\Gamma }d^{2}w\left( \partial \varphi 
\bar{\partial}\varphi +b^{2}e^{2b\varphi }\right) \right) \right] \times \\
& \times \exp \left[ -\frac{i\xi }{4\pi }\left( \frac{u}{|u|}\right)
^{m}\int_{\partial \Gamma }\ d\tau \left( \frac{|\tau -z|}{\tau -z}\,\frac{%
|\tau -\bar{z}|}{\tau -\bar{z}}\,\frac{|\rho (z)|}{\rho (z)}\right)
^{m}e^{b\varphi }\right] \,e^{2b(j+1)}\,e^{\frac{m}{b}\varphi (z)}\times \\
& \times |\mu |^{2m(j+1)}\,|z-\bar{z}|^{2m(j+1)}\,|u|^{-2m(j+1)}|\rho
(z)|^{2h_{j}+\frac{m^{2}}{2b^{2}}}
\end{split}%
\end{equation}

Then, taking into account (\ref{L42}), 
\begin{equation}
e^{2b(j+1)\varphi (z,\bar{z})}\,e^{\frac{m}{b}\varphi (z,\bar{z})}\ =\
e^{\left( 2b(j+1)+\frac{m}{b}\right) \varphi (z,\bar{z})}\,|\rho
(z)|^{2m(j+1)},
\end{equation}%
what amounts to extract the pole in the coincidence limit of the two
operators $e^{2b(j+1)\varphi (z,\bar{z})}$ and $e^{\frac{m}{b}\varphi (w,%
\overline{w})}$. Then we find%
\begin{equation}
\begin{split}
\Omega _{j}^{(m,b)}(z)& =\ \delta ^{2}(\mu +\bar{\mu})|\mu
|^{mb^{-2}(b^{2}+1-m)}(2y)^{mb^{-2}(b^{2}+1-m/2)}\int \mathcal{D}\varphi
\exp \left[ -\frac{1}{2\pi }\int_{\Gamma }d^{2}w\left( \partial \varphi \bar{%
\partial}\varphi +b^{2}\,e^{2b\varphi }\right) \right] \times \\
& \times \exp \left[ -\frac{i\xi }{4\pi }(\text{sgn}(\text{Im}\mu
))^{m}\int_{\partial \Gamma }d\tau \,e^{b\varphi }\right] |\rho
(z)|^{2h_{j}+m(b^{2}+1-m)/b^{2}}\,e^{\left( 2b(j+1)+m/b\right) \varphi (z,%
\bar{z})}
\end{split}
\label{L50}
\end{equation}%
where \text{sgn}$($\text{Im}$\mu )$ is the sign of the imaginary part of $%
\mu $. Notice that this expression is well defined for $m\in \mathbb{Z}$, as
in that case the quantity $($\text{sgn}$($\text{Im}$\mu ))^{m}$ is real.
Remarkably, the expression on the right hand side of (\ref{L50}) corresponds
to the disk one-point function in Liouville field theory multiplied by a
factor $\delta ^{2}(\mu +\bar{\mu})|\mu
|^{mb^{-2}(b^{2}+1-m)}(2y)^{mb^{-2}(b^{2}+1-m/2)}$, provided one agrees on
identifying the Liouville parameters $\alpha ,$ $b,$ $\mu _{L}$ and $\mu
_{B} $ as follows%
\begin{eqnarray}
\mu _{L} &=&\frac{b^{2}}{2\pi }=\frac{\lambda }{2\pi }, \\
\mu _{B} &=&\frac{i\xi }{4\pi }(\text{sgn}(\text{Im}\mu ))^{m}=\frac{\lambda
_{B}}{4\pi }, \\
\alpha &=&b(j+1)+\frac{m}{2b}, \\
h_{\alpha } &=&\alpha (Q_{L}-\alpha )\ =h_{j}+\frac{m}{2b^{2}}(b^{2}+1-\frac{%
m}{2}),
\end{eqnarray}%
where we are using the standard notation;\ see for instance \cite%
{teschner,Yu}. This is to say 
\begin{equation}
\bigg\langle\Phi _{j}(\mu |z)\bigg\rangle_{\mathcal{T}_{m,b}}=\delta
^{2}(\mu +\bar{\mu})|\mu |^{m(1+\frac{1-m}{b^{2}})}\,(2y)^{m(1+\frac{1}{b^{2}%
}-\frac{m}{b^{2}})}\bigg\langle V_{\alpha }(z)\bigg\rangle_{\text{Liouville}}
\label{54}
\end{equation}%
where $V_{\alpha }(z)=e^{2\alpha \varphi (z,\bar{z})}$. This trick, which is
exactly the one used in \cite{ribault2} to solve the case $m=1$, leads us to
obtain the explicit expression for $\Omega _{j}^{(m,b)}$ in terms of the
Liouville one-point function $\Omega _{j+m/2b^{2}}^{(m=0,b)}$. In fact, the
expression for the disk one-point function of a bulk operator in Liouville
theory is actually known \cite{FZZ, teschner}; it reads 
\begin{equation}
\bigg\langle V_{\alpha }(z)\bigg\rangle_{\text{Liouville}}\ =\ |z-\bar{z}%
|^{-2h_{\alpha }}\frac{2}{b}\Omega _{0}\left( \pi \mu \frac{\Gamma (b^{2})}{%
\Gamma (1-b^{2})}\right) ^{\frac{Q-2\alpha }{2b}}\,\cosh (2\pi s(2\alpha
-Q))\,\Gamma (2b\alpha -b^{2})\,\Gamma \left( \frac{2\alpha -Q}{b}\right)
\label{L1}
\end{equation}%
where the parameter $s$ obeys 
\begin{equation}
\cosh (2\pi bs)\ =\ \frac{\mu _{B}}{\sqrt{\mu _{L}}}\sqrt{\sin (\pi b^{2})}
\label{L3}
\end{equation}%
and where $\Omega _{0}$ is an irrelevant overall factor we determine below
(see Section 5).

Thus, by following the trick in \cite{ribault2} (see also \cite{Ribault2,
hosomichiribault, hosomichi}), we managed to calculate the expectation value 
$\Omega _{j}^{(m,b)}$ of a bulk operator in the theory (\ref{Sbulk}) on the
disk by reducing such calculation to that of the observable $\Omega _{j=%
\frac{\alpha }{b}-\frac{m}{2b^{2}}-1}^{(m=0,b)}$ in Liouville field theory.
The final result for $m\in \mathbb{Z}$ reads 
\begin{eqnarray}
\Omega _{j}^{(m,b)}\  &=&\ \frac{2}{b}\Omega _{0}\,\delta (\mu +\bar{\mu}%
)|\mu |^{m(1+\frac{1-m}{b^{2}})}\,|z-\overline{z}|^{-2h_{j}}\left( \pi \,%
\frac{\Gamma (1-b^{2})}{\Gamma (b^{2}+1)}\right) ^{j+\frac{1}{2}-\frac{1-m}{%
2b^{2}}}\Gamma \left( 2j+1-\frac{1-m}{b^{2}}\right) \times   \notag \\
&&\times \Gamma \left( b^{2}(2j+1)+m\right) \,\cosh \left( (r-i\frac{m\pi }{2%
}(\text{sign}(\text{Im}\mu )))\left( 2j+1-\frac{1-m}{b^{2}}\right) \right) 
\label{nueve}
\end{eqnarray}%
with $h_{j}=-b^{2}j(j+1)+(j+1)(1-m)$ and where we introduced the parameter 
\begin{equation}
r=2\pi bs+i\frac{m\pi }{2}\text{sign}(\text{Im}\mu ).  \label{segundopar}
\end{equation}%
This boundary parameter $r$ is analogous to the one introduced in \cite%
{ponsotschomerusteschner} for the WZNW\ theory $m=1$. Here we are involved
with the case of $m\in \mathbb{Z}$, for which the Lagrangian representation (%
\ref{S}) makes sense, and then the boundary parameter $r$ is related to $\xi 
$ as follows
\begin{equation}
\cosh (r-i\frac{m\pi }{2}\text{sign}(\text{Im}\mu ))\ =\ i\,\xi \,(\text{sgn}%
(\text{Im}\mu ))^{m}\,\sqrt{\frac{\sin (\pi b^{2})}{8\pi b^{2}}}.
\label{siete}
\end{equation}%

Again, as it happens in Liouville theory, relation (\ref{siete}) exhibits a
symmetry under the shift $r\rightarrow r+2\pi i$, while definition (\ref%
{asdefined}) does not. Consequently, it is (\ref{asdefined}) (and not (\ref{siete}))
the expression that has be considered as the definition of $r$ in
terms of the Liouville parameter $s$.

We already mentioned in the Introduction that expression (\ref{54}) fulfils
several non-trivial consistency checks; for example, the reflection equation
(\ref{torbe2}). Actually, (\ref{54}), together with (\ref{segundopar}),
could be regarded as the conjecture for the bulk one-point function of the
CFTs are thought to exist.

In the next section we will reobtain the result (\ref{nueve}) using the free
field calculation. As the path integral derivation, the free field
derivation will be also valid in the case of $m$ being an integer number.

\section{Free field computation}

In this section, we compute the bulk one-point function in a different way.
We will consider the free field approach, which has proven to be a useful
method to calculate correlation functions in this type of non-rational
models on the sphere \cite{GN3, Becker, Ponjas}.

We consider Newman boundary conditions for the fields; namely on the
boundary $\partial \Gamma $ we demand $\partial \phi -\bar{\partial}\phi $, $%
\beta +\bar{\beta}$ and $\gamma +\bar{\gamma}$ to vanish. This is consistent
with the boundary conditions (\ref{tres})-(\ref{Dos}) in the perturbative
free field approximation (where $\xi =0$ is considered to compute the
correlators and the boundary interaction term is realized by additional
insertions perturbatively). For such boundary conditions on the geometry of
the disk, the non-vanishing correlators turn out to be 
\begin{eqnarray}
\big\langle\phi (z,\bar{z})\phi (w,\bar{w})\big\rangle &=&-\ln |z-w||z-\bar{w%
}|\ , \\
\big\langle\beta (z)\gamma (w)\big\rangle &=&-\frac{1}{z-w},\quad \big\langle%
\bar{\beta}(\bar{z})\bar{\gamma}(\bar{w})\big\rangle=-\frac{1}{\bar{z}-\bar{w%
}}, \\
\big\langle\bar{\beta}(\bar{z})\gamma (w)\big\rangle &=&\frac{1}{\bar{z}-w}%
,\quad \big\langle\beta (z)\bar{\gamma}(\bar{w})\big\rangle=\frac{1}{z-\bar{w%
}}.
\end{eqnarray}%
where we see the mixing between holomorphic and anti-holomorphic modes due
to the presence of the boundary conditions.

Now, we have to verify that the boundary term we added to the action
actually corresponds to a theory that preserves the symmetry generated by $%
J^{-}(z)$, $J^{3}(z)$ and $T(z)$. Checking this at the first order in $%
\lambda _{B}$ amounts to check that the following expectation values vanish, 
\begin{eqnarray}
\left\langle \left( J^{-}(z)+\bar{J}^{-}(\bar{z})\right) \int_{\partial
\Gamma }d\tau \beta ^{m}(\tau )e^{b\phi (\tau )}\ldots \right\rangle _{|z=%
\overline{z}} &=&0,  \label{80000} \\
\left\langle \left( J^{3}(z)-\bar{J}^{3}(\bar{z})\right) \int_{\partial
\Gamma }d\tau \beta ^{m}(\tau )e^{b\phi (\tau )}\ldots \right\rangle _{|z=%
\overline{z}} &=&0.  \label{800}
\end{eqnarray}

Once again, this is completely analogous to the analysis done in \cite%
{ribault2} for $m=1$. However, in contrast to the case of the WZNW theory,
where the condition $J^{+}(z)+\bar{J}^{+}(\bar{z})=0$ in known to hold as
well \cite{ribault2}, in the case of generic $m$ we have to impose%
\begin{equation}
\left\langle \left( T(z)-\bar{T}(\bar{z})\right) \int_{\partial \Gamma
}d\tau \beta ^{m}(\tau )e^{b\phi (\tau )}\ldots \right\rangle _{|z=\overline{%
z}}=0  \label{80}
\end{equation}%
explicitly. To verify that conditions (\ref{80000})-(\ref{80}) are obeyed,
first we compute the following operator product expansions (OPEs) 
\begin{eqnarray}
J^{-}(z)\beta ^{m}(\tau )e^{b\phi (\tau )}\ &\sim &\ 0 \\
\bar{J}^{-}(\bar{z})\beta ^{m}(\tau )e^{b\phi (\tau )}\ &\sim &\ 0
\end{eqnarray}%
and%
\begin{eqnarray}
J^{3}(z)\beta ^{m}(\tau )e^{b\phi (\tau )}\ &\sim &\ \ 0 \\
\bar{J}^{3}(\bar{z})\beta ^{m}(\tau )e^{b\phi (\tau )}\ &\sim &\ \ 0
\end{eqnarray}%
where we considered that in the boundary $\bar{\beta}(x)=-\beta ({x)}$, and
where the symbol $\sim 0$ means that the singular terms vanish when
evaluating in the boundary.

Next, we have to impose condition (\ref{80}). To do so, we have to compute
the OPE 
\begin{equation}
T(z)\beta ^{m}(\tau )e^{b\phi (\tau )}\ \sim \ \partial _{\tau }\left( \frac{%
\beta ^{m}(\tau )e^{b\phi (\tau )}}{z-\tau }\right) -\frac{ib\partial
_{\sigma }\phi (\tau )\beta ^{m}(\tau )e^{b\phi (\tau )}}{z-\tau }+...
\label{85}
\end{equation}%
where, again, the symbol $\sim $ means that the equivalence holds up to
regular terms and exact differentials when evaluating in the boundary.
Analogously, we have the anti-holomorphic part 
\begin{equation}
\bar{T}(\bar{z})\beta ^{m}(\tau )e^{b\phi (\tau )}\ \sim \partial _{\tau
}\left( \frac{\beta ^{m}(\tau )e^{b\phi (\tau )}}{\bar{z}-\tau }\right) -%
\frac{ib\partial _{\sigma }\phi (\tau )\beta ^{m}(\tau )e^{b\phi (\tau )}}{%
\bar{z}-\tau }\ +...  \label{86}
\end{equation}

In the boundary we find that (\ref{85})\ and (\ref{86}) contribute with the
same piece and then we verify that (\ref{80}) is actually satisfied.

Then, we can proceed and compute the bulk one-point function using the free
field approach. This observable is given by 
\begin{equation}
\begin{split}
\Omega _{j}^{(m,b)}\ & =\ \int \mathcal{D}\phi \mathcal{D}^{2}\beta \mathcal{%
D}^{2}\gamma \,\exp \bigg[-\frac{1}{2\pi }\int\nolimits_{\Gamma }\partial
\phi \bar{\partial}\phi +\frac{1}{2\pi }\int_{\Gamma }\gamma \bar{\partial}%
\beta +\frac{1}{2\pi }\int_{\Gamma }\bar{\gamma}\partial \bar{\beta}-\frac{%
Q_{m}}{8\pi }\int_{\Gamma }g^{1/2}R\phi - \\
& -\frac{b^{2}}{2\pi }\int_{\Gamma }(-\beta \bar{\beta})^{m}e^{2b\phi }-%
\frac{Q_{m}}{2\pi }\int_{\partial \Gamma }g^{1/4}K\phi +\frac{i\xi }{4\pi }%
\int_{\partial \Gamma }\beta ^{m}e^{b\phi }\bigg]|\mu |^{2m(j+1)}\,e^{\mu
\gamma (z)-\bar{\mu}\bar{\gamma}(\bar{z})}\,e^{2b(j+1)\phi (z,\bar{z})}
\end{split}%
\end{equation}%
with $Q_{m}=b+\frac{1-m}{b}$. The notation we use is such that $z=x+iy$ and $%
w=\tau +i\sigma $.

Splitting the field $\phi $ in its zero-mode $\phi _{0}$ and its
fluctuations $\phi ^{\prime }=\phi -\phi _{0}$, and recalling that the
Gauss-Bonnet contribution gives 
\begin{equation}
\frac{1}{8\pi }\int_{\Gamma }g^{1/2}R\text{ }Q_{m}\phi _{0}+\frac{1}{2\pi }%
\int_{\partial \Gamma }g^{1/4}K\text{ }Q_{m}\phi _{0}\ =\ Q_{m}\phi _{0},
\end{equation}%
we find 
\begin{equation}
\begin{split}
\Omega _{j}^{(m,b)}\ & =\ \int \mathcal{D}\phi ^{\prime }d\phi _{0}\mathcal{D%
}^{2}\beta \mathcal{D}^{2}\gamma \,e^{-S_{|\lambda =0}^{\prime }}\exp \left[ 
\frac{-b^{2}}{2\pi }e^{2b\phi _{0}}\int_{\Gamma }(-\beta \bar{\beta}%
)^{m}e^{2b\phi ^{\prime }}\right] \exp \left[ \frac{i\xi }{4\pi }e^{b\phi
_{0}}\int_{\partial \Gamma }\beta ^{m}e^{b\phi ^{\prime }}\right] \times \\
& \times e^{2b(j+1)\phi _{0}}\,e^{-Q_{m}\phi _{0}}\,|\mu |^{2m(j+1)}\,e^{\mu
\gamma (z)-\bar{\mu}\bar{\gamma}(\bar{z})}\,e^{2b(j+1)\phi ^{\prime }(z,\bar{%
z})}
\end{split}%
\end{equation}%
where $S_{|\lambda =0}^{\prime }$ means the free action, i.e. the action (%
\ref{S}) with $\lambda =\lambda _{B}=0$ evaluated on the field fluctuations $%
\phi ^{\prime }$.

Then, we have to integrate over the zero-mode $\phi _{0}$. The interaction
terms in the actions give the following contribution to the integrand 
\begin{equation}
\exp \left( \frac{-b^{2}}{2\pi }e^{2b\phi _{0}}\int_{\Gamma }(-\beta \bar{%
\beta})^{m}e^{2b\phi ^{\prime }}+\frac{i\xi }{4\pi }e^{b\phi
_{0}}\int_{\partial \Gamma }\beta ^{m}e^{b\phi ^{\prime }}\right) ,
\label{integrand}
\end{equation}%
while the vertex operator itself contributes with an exponential $%
e^{(2b(j+1)-Q_{m})\phi _{0}}$. There is also a contribution from the
background charge. For the disk geometry, both bulk and boundary screening
operators are present. Standard techniques in the free field calculation
yield the following expression for the residues of resonant correlators, 
\begin{equation}
\begin{split}
\underset{2j+1+\frac{m-1}{b^{2}}=-n}{\text{Res}~\Omega _{j}^{(m,b)}}& =\ 
\frac{1}{2b}\,|\mu |^{2m(j+1)}\sum_{\underset{2p+l=n}{p,l=0}}^{\infty }\frac{%
1}{p!l!}\prod_{i=1}^{\infty }\int_{\Gamma }d^{2}w_{i}\prod_{k=1}^{\infty
}\int_{\partial \Gamma }d\tau _{k}\bigg\langle e^{\mu \gamma (z)-\bar{\mu}%
\bar{\gamma}(\bar{z})}\times \\
& \times e^{2b(j+1)\phi (z,\bar{z})}\prod_{i=1}^{p}\frac{(-b^{2})}{2\pi }%
(-\beta \bar{\beta})^{m}\,e^{2b\phi (w_{i},\bar{w}_{i})}\prod_{k=1}^{l}\frac{%
i\xi }{4\pi }\beta ^{m}\,e^{b\phi (x_{k})}\bigg\rangle
\end{split}
\label{MS}
\end{equation}%
where, as mentioned, now both bulk and boundary screening operators, $%
\int_{\Gamma }(-\beta \bar{\beta})^{m}e^{2b\phi }$ and $\int_{\partial
\Gamma }\beta ^{m}e^{b\phi }$, appear. As shown above, the integration over
the zero-mode $\phi _{0}$ gives a charge conservation condition that demands
to insert a precise amount of screening operators for the correlators not to
vanish; $p$ of these screening operators are to be inserted in the bulk,
while $l$ of them in the boundary, with $2p+l=n$. The precise relation is 
\begin{equation}
2b(j+1)+2\,b\,p+b\,l\ =\ Q_{m}\ =\ b+\frac{1-m}{b};  \label{MS2}
\end{equation}%
that is $n\ =\ 2\,p+l\ =-2j-1+(1-m)/b^{2}$.

The correlator then factorizes out in two parts: the part that depends on $%
\phi ,$ and the contribution of the $\gamma $-$\beta $ ghost system. The
Coulomb gas calculation of the $\phi $ contribution yields 
\begin{equation}
\begin{split}
\bigg\langle e^{2(j+1)\phi (iy)}\prod_{i=1}^{p}& e^{2b\phi (w_{i},\bar{w}%
_{i})}\prod_{k=1}^{l}e^{b\phi (\tau _{k})}\bigg\rangle\ =\ \left[
\prod_{k=1}^{l}(y^{2}+\tau _{k}^{2})\prod_{i=1}^{p}|y^{2}+w_{i}^{2}|^{2}%
\right] ^{-2b^{2}(j+1)}\times \\
& \times |2y|^{-2b^{2}(j+1)^{2}}\left[ \prod_{i,k}|w_{i}-x_{k}|^{2}%
\prod_{i<i^{\prime }}|w_{i}-w_{i^{\prime }}|^{2}\prod_{i,i^{\prime }}|w_{i}-%
\bar{w}_{i^{\prime }}|\prod_{k<k^{\prime }}|x_{k}-x_{k^{\prime }}|\right]
^{-2b^{2}}
\end{split}%
\end{equation}

On the other hand, for working out the ghost contribution $\gamma $-$\beta $
it is convenient first to consider the OPE 
\begin{equation}
\bigg\langle e^{\mu \gamma (z)-\bar{\mu}\bar{\gamma}(\bar{z})}\beta (w)%
\bigg\rangle\ =\ \frac{\mu }{w-z}+\frac{\bar{\mu}}{w-\bar{z}}\ =\ \frac{\mu
\,(z-\bar{z})}{(w-z)(w-\bar{z})}.
\end{equation}%
This implies that the $\gamma $-$\beta $ correlator takes the form 
\begin{equation}
\left\langle e^{\mu \gamma (z)-\bar{\mu}\bar{\gamma}(\bar{z}%
)}\prod_{i=1}^{p}\left( -\beta (w_{i})\bar{\beta}(\bar{w}_{i})\right)
^{m}\right\rangle \ =\ \mu ^{2pm}\,(2iy)^{2pm}\,\prod_{i=1}^{p}\frac{1}{%
|y^{2}+w_{i}^{2}|^{2m}}
\end{equation}%
and 
\begin{equation}
\left\langle e^{\mu \gamma (z)-\bar{\mu}\bar{\gamma}(\bar{z}%
)}\prod_{k=1}^{l}\beta ^{m}(\tau _{k})\right\rangle \ =\ \mu
^{lm}\,(2iy)^{lm}\prod_{k=1}^{l}\frac{1}{(y^{2}+\tau _{k})^{m}}.
\end{equation}

The full correlator is then given by 
\begin{equation}
\begin{split}
& \left\langle e^{\mu \gamma (iy)-\bar{\mu}\bar{\gamma}(-iy)}\prod_{i=1}^{p}%
\frac{(-b^{2})}{2\pi }(-\beta (w_{i})\bar{\beta}(\bar{w}_{i}))^{m}%
\prod_{k=1}^{l}\frac{i\xi }{4\pi }\beta (\tau _{k})^{m}\right\rangle \ = \\
& =\ 2\pi \,\delta (\mu +\bar{\mu})\,\left( \frac{b^{2}}{2\pi }\right)
^{p}\left( \frac{i\xi }{4\pi }(-\text{sgn}(\text{Im}\mu ))^{m}\right)
^{l}|2u|^{nm}\,|\mu |^{nm}\prod_{i=1}^{p}\frac{1}{|y^{2}+w_{i}^{2}|^{2m}}%
\prod_{k=1}^{l}\frac{1}{(y^{2}+\tau _{k}^{2})^{m}}
\end{split}%
\end{equation}


Returning to the form of the resonant correlators, for which $p$ and $l$ are
integer numbers, the residues of these observables take the form 
\begin{equation}
\begin{split}
& \underset{2j+1+\frac{m-1}{b^{2}}=-n}{\text{Res}~\Omega ^{(m,b)}}\ =\ \frac{%
\pi }{b}\,\sum_{p=0}^{\infty }\sum\limits_{l=0}^{\infty }\frac{1}{p!l!}%
\delta _{2p+l-n,0}(-1)^{p}\left( \frac{b^{2}}{2\pi }\right) ^{p}\left( \frac{%
i\xi }{4\pi }(-\text{sgn}(\text{Im}\mu ))^{m}\right) ^{l}\times \\
& \times \int \prod_{i=1}^{p}\frac{d^{2}w_{i}}{%
|y^{2}+w_{i}^{2}|^{4b^{2}(j+1)+2m}}\int \prod_{k=1}^{l}\frac{dx_{k}}{%
(y^{2}+x_{k}^{2})^{b^{2}(j+1)+m}}\bigg[\prod_{i,k}|w_{i}-x_{k}|^{2}%
\prod_{i<i^{\prime }}|w_{i}-w_{i^{\prime }}|^{2}\times \\
& \times \prod_{i,i^{\prime }}|w_{i}-\bar{w}_{i^{\prime
}}|\prod_{k<k^{\prime }}|x_{k}-x_{k^{\prime }}|\bigg]^{-2b^{2}}\,\delta (\mu
+\bar{\mu})|\mu |^{m\left( 1+\frac{1-m}{b^{2}}\right)
}\,|2y|^{-2b^{2}(j+1)^{2}+nm}.
\end{split}%
\end{equation}

To solve this we use the following integral formula (see Ref. \cite{ribault2}
for the computation in the case $m=1$) 
\begin{equation}
\begin{split}
Y_{n,p}(a)\,& =\,\frac{1}{p!(n-2p)!}\int \prod_{i=1}^{p}\frac{d^{2}w_{i}}{%
|y^{2}+w_{i}^{2}|^{2a}}\int \prod_{k=1}^{l}\frac{dx_{k}}{%
(y^{2}+x_{k}^{2})^{a}}\bigg[\prod_{i,k}|w_{i}-x_{k}|^{2}\prod_{i<i^{\prime
}}|w_{i}-w_{i^{\prime }}|^{2}\times \\
& \times \prod_{i,i^{\prime }}|w_{i}-\bar{w}_{i^{\prime
}}|\prod_{k<k^{\prime }}|x_{k}-x_{k^{\prime }}|\bigg]^{-2b^{2}}
\end{split}%
\end{equation}%
with $a=2b^{2}(j+1)+m=1+b^{2}-b^{2}n$. The solution of this multiple
Selberg-type integral is given by 
\begin{equation}
Y_{n,p}(a)\ =\ |z-\bar{z}|^{n\left( 1-2a-(n-1)b^{2}\right) }\left( \frac{%
2\pi }{\Gamma (1-b^{2})}\right) ^{n}\frac{2^{-2p}}{n!(\sin (\pi b^{2}))^{p}}%
\,I_{n}(a)\,J_{n,p}(a),
\end{equation}%
where 
\begin{equation}
I_{n}(a)\ =\ \prod_{i=0}^{n-1}\frac{\Gamma \left( 1-(i-1)b^{2}\right) \Gamma
\left( 2a-1+(n-1+i)b^{2}\right) }{\Gamma ^{2}\left( a+ib^{2}\right) }
\label{MS3}
\end{equation}%
and 
\begin{eqnarray}
J_{n,p}(a) &=&\sum\limits_{i=0}^{p}(-1)^{i}\frac{\Gamma (n-p-i+1)}{\Gamma
(p-i+1)\Gamma (n-2p+1)}\frac{\sin (\pi b^{2}(n-2i+1))}{\sin (\pi
b^{2}(n-i+1))}\times  \notag \\
&&\times \prod\limits_{r=0}^{i-1}\frac{\sin (\pi b^{2}(n-r))\sin (\pi a+\pi
b^{2}(n-r))}{\sin (\pi b^{2}(r+1))\sin (\pi a+\pi b^{2}r)}.
\end{eqnarray}%
For the case of our interest, $a=1+b^{2}(1-n)$, and the function $I_{n}(a)$
simplifies substantially, taking the value $I_{n}(a)\ =\ \Gamma \left(
1-b^{2}n\right) $. The sum over $J_{n,p}(a)$ also simplifies notably; see
below.

Then, we find the expression 
\begin{equation}
\begin{split}
\underset{2j+1+\frac{m-1}{b^{2}}=-n}{\text{Res}~\Omega ^{(m,b)}}& =\ \frac{%
\pi }{b}\,\delta (\mu +\bar{\mu})|\mu |^{m\left( 1+\frac{1-m}{b^{2}}\right)
}\,|z-\bar{z}|^{-2h_{j}}\left( \frac{2\pi }{\Gamma (1-b^{2})}\right) ^{n}%
\frac{\Gamma (1-b^{2}n)}{n!}\times \\
& \times \sum_{\underset{2p+l=n}{p,l=0}}^{\infty }(-1)^{p}\left( \frac{b^{2}%
}{2\pi }\right) ^{p}\left( \frac{i\xi }{4\pi }\left( -\text{sgn}(\text{Im}%
\mu )\right) ^{m}\right) ^{l}\frac{2^{-2p}}{(\sin (\pi b^{2}))^{p}}%
\,J_{n,p}(a)
\end{split}
\label{ocho}
\end{equation}%
with $h_{j}\,=\,b^{2}j(j+1)-(j+1)(1-m).$

We may try to simplify the expression above further. Following \cite%
{ribault2}, we find 
\begin{equation}
\sum_{p=0}^{|n/2|}(-1)^{p}\,(2\cosh (2\pi
bs))^{n-2p}\,J_{n,p}(1+b^{2}(1-n))\ =\ \cosh (2\pi nbs),
\end{equation}%
where $2\pi bs=r-i({m \pi }/{2})\text{sgn}(\text{Im}\mu )$ and where we
replaced $a=1+b^{2}-b^{2}n$. According to the notation introduced in
(\ref{siete}), we write 
\begin{equation}
\xi \ =\ i(-\text{sgn}(\text{Im}\mu ))^{m}\,\sqrt{\frac{8\pi b^{2}}{\sin
(\pi b^{2})}}\,\cosh (r-i\frac{m\pi }{2}\text{sgn}(\text{Im}\mu )),
\end{equation}%
and this yields 
\begin{equation}
\left( \frac{i\xi }{4\pi }\left( -\text{sgn}(\text{Im}\mu )\right)
^{m}\right) ^{n-2p}\ =\ \left( -\frac{1}{4\pi }\sqrt{\frac{8\pi b^{2}}{\sin
(\pi b^{2})}}\right) ^{n-2p}(\cosh (r-i\frac{m\pi }{2}\text{sgn}(\text{Im}%
\mu )))^{n-2p}.
\end{equation}

Replacing this into the sum in (\ref{ocho}) we obtain 
\begin{eqnarray}
\sum\limits_{p=0}^{\infty }\sum\limits_{l=0}^{\infty }\delta
_{2p+l-n,0}(-1)^{p}\left( \frac{b^{2}}{2\pi }\right) ^{p}\left( \frac{i\xi }{%
4\pi }\left( -\text{sgn}(\text{Im}\mu )\right) ^{m}\right) ^{l}\frac{2^{-2p}%
}{(\sin (\pi b^{2}))^{p}}J_{n,p}(a)\  =  \label{MS5} \\
=\ (-1)^{n}\,2^{-\frac{3}{2}n}\,\pi ^{-n}\left( \Gamma (1-b^{2})\Gamma
(1+b^{2})\right) ^{\frac{n}{2}}\cosh (n(r-i\frac{m\pi }{2}\text{sgn}(\text{Im%
}\mu ))),  \notag
\end{eqnarray}%
which can be proven using, in particular, the relation $\pi /\sin (\pi
b^{2})=\Gamma (1-b^{2})\Gamma (b^{2})$.

Putting all the pieces together, the final result for the resonant
correlators reads 
\begin{equation}
\begin{split}
\underset{2j+1+\frac{m-1}{b^{2}}=-n}{\text{Res}~\Omega ^{(m,b)}}& =\ 2^{-%
\frac{n}{2}}\,\frac{\pi }{b}\,\delta (\mu +\bar{\mu})|\mu |^{m\left( 1+\frac{%
1-m}{b^{2}}\right) }\,|z-\bar{z}|^{-2h_{j}}\,\frac{(-1)^{n}}{n!}\left( \frac{%
\Gamma (1+b^{2})}{\Gamma (1-b^{2})}\right) ^{\frac{n}{2}}\times  \\
& \times \Gamma (1-b^{2}n)\,\cosh (n(r-i\frac{m\pi }{2}\text{sgn}(\text{Im}%
\mu )))
\end{split}
\label{144}
\end{equation}

This can be rewritten by recalling that for $n\in \mathbb{Z}_{\geq 0}$ it
holds%
\begin{equation}
\lim_{\varepsilon \rightarrow 0}\varepsilon \Gamma (\varepsilon -n)=(-1)^{n}%
\frac{1}{\Gamma (n+1)},
\end{equation}%
which allows to replace the factor $(-1)^{n}/n!$ in (\ref{144}) as follows%
\begin{equation}
\frac{(-1)^{n}}{\Gamma (n+1)}\rightarrow \Gamma (2j+1+(m-1)b^{-2}).
\end{equation}%
were we used $(1-m)b^{-2}-2j-1=n.$ Written in terms of $j$, $b$, and $m$,
the final result reads 
\begin{equation}
\begin{split}
\Omega _{j}^{(m,b)}\ & =\left( \frac{\pi }{2}\right) ^{\frac{n}{2}}\frac{\pi 
}{b}\ \,\delta (\mu +\bar{\mu})|\mu |^{m(1+\frac{1-m}{b^{2}})}\,|z-\bar{z}%
|^{-2h_{j}}\left( \pi \,\frac{\Gamma (1-b^{2})}{\Gamma (1+b^{2})}\right)
^{(2j+1+(m-1)b^{-2})/2}\times  \\
& \times \Gamma \left( 2j+1-\frac{1-m}{b^{2}}\right) \,\Gamma \left(
b^{2}(2j+1)+m\right) \,\cosh \left( (r-i\frac{m\pi }{2}\text{sgn}(\text{Im}%
\mu ))\left( 2j+1-\frac{1-m}{b^{2}}\right) \right) 
\end{split}
\label{93}
\end{equation}%
with $h_{j}=-b^{2}j(j+1)+(j+1)(1-m)$ and $\cosh (r-i\frac{m\pi }{2}\text{sgn}%
(\text{Im}\mu ))=i\xi ($sgn(Im$\mu ))^{m}\sqrt{\sin (\pi b^{2})/8\pi b^{2}}$%
. And we see that expression (\ref{93}) coincides with the path integral
result (\ref{nueve}) if we chose the numerical global coefficient in (\ref%
{nueve}) to be $\Omega _{0}=\left( \pi /2\right) ^{n/2}.$ Therefore, we find
exact agreement between the path integral calculation and the free field
calculation. This manifestly shows that free field approach turns out to be
a useful method to find the general expression of correlation functions in
non-rational CFTs. The path integral method actually reduced the problem to
that of computing an observable of Liouville theory, which may be done by
different methods, e.g. the bootstrap method. The relation existing between
correlators of the conformal theories $\mathcal{T}_{m,b}$ defined by the
action (\ref{S}) and correlators of Liouville theory is a generalization of
the so called $H_{3}^{+}$ WZNW-Liouville correspondence \cite{Stoyanovsky,
ribaultteschner, Ribault1, schomerus}. The free field approach, on the other
hand, gave a quite direct computation of the disk one-point function.

\section{Analysis of the one-point function}

Let us first analyze some special cases of the general result (\ref{93}).
The first particular example we may consider is clearly the theory for $m=0$%
, namely $\mathcal{T}_{m=0,b}$. In this case, one-point function (\ref{93})
trivially reduces to Liouville disk one-point function (\ref{L1}).
Normalized states $\left\vert \alpha \right\rangle $ of the theory are
created by the action of exponential vertex operators on the vacuum, $%
\lim_{z\rightarrow 0}e^{2\alpha \varphi (z)}\left\vert 0\right\rangle
=\left\vert \alpha \right\rangle $, having momentum $\alpha =Q+iP$, with $%
P\in \mathbb{R}$ and $Q=b+b^{-1}$; see for instance \cite{Teschner, Yu,
Teschnerexponential}. Then, Liouville one-point function reads 
\begin{equation}
\bigg\langle V_{\alpha }(z)\bigg\rangle_{\text{Liouville}}\ =\ |z-\bar{z}%
|^{-2h_{\alpha }}\left( \pi \frac{\Gamma (b^{2})}{\Gamma (1-b^{2})}\right) ^{%
\frac{-iP}{b}}\,\frac{\cos (4\pi sP)}{iP}\,\Gamma (1+2ibP)\,\Gamma \left( 1+%
\frac{2iP}{b}\right)
\end{equation}%
where $s$ given by $\cosh (2\pi bs)=\sqrt{\sin (\pi b^{2})\mu _{B}^{2}/\mu
_{L}}$, and where we have fixed the Liouville "cosmological constant" $\mu
_{L}$ to a specific value.

Next, we have the case $m=1$, which corresponds to the $H_{+}^{3}$ WZNW\
theory with level $k$, $\mathcal{T}_{m=1,b=(k-2)^{-1/2}}$. In this case, the
observable $\Omega _{j}^{(m=1,b=(k-2)^{-1/2})}=\left\langle \Phi _{j}(\mu
|z)\right\rangle _{\text{WZNW}}$ represents AdS$_{2}$ branes in Euclidean AdS%
$_{3}$ space. This is 
\begin{eqnarray}
\bigg\langle\Phi _{j}(\mu |z)\bigg\rangle_{\text{WZNW}} &=&\pi \sqrt{k-2}%
\,\delta (\mu +\bar{\mu})\text{ }|\mu |\text{ }\,|z-\overline{z}%
|^{-2h_{j}}\left( \pi \,\frac{\Gamma (1+\frac{1}{k-2})}{\Gamma (1-\frac{1}{%
k-2})}\right) ^{ip/2}\times  \notag \\
&&\times \Gamma \left( ip\right) \,\Gamma \left( 1+\frac{ip}{k-2}\right)
\,\cosh \left( irp+\frac{\pi }{2}p(\text{sign}(\text{Im}\mu ))\right) .
\label{999}
\end{eqnarray}%
where $j=-1/2+ip$ belongs to the continuous representation, with $p\in 
\mathbb{R}$, and where the parameter $r$ is that introduced in Ref. \cite%
{ponsotschomerusteschner}. The string coupling constant $g_{s}^{2}=e^{-2%
\left\langle \vartheta \right\rangle \chi }=e^{-\chi /\sqrt{2k-4}}$ was also
fixed to a specific value to absorb $\Omega _{0}$ (here, $\left\langle
\vartheta \right\rangle $ represents the expectation value of the dilaton
field, and $\chi $ is the Euler characteristic of the worldsheet manifold.)
In terms of fields $\phi $, $\gamma $ and $\gamma $ we used to describe the
theory (\ref{Sbulk}), AdS$_{3}$ metric is written in Poincar\'{e}
coordinates as $ds^{2}=l^{2}\left( d\phi ^{2}+e^{2\phi }d\gamma d\overline{%
\gamma }\right) $, where $l$ is the "radius" of the space. When formulating
string theory on this background, the level of the WZNW theory relates to
the string tension as follows $k=l^{2}/\alpha ^{\prime }$. Branes in
Lorentzian and Euclidean AdS$_{3}$ space were extensively studied in the
literature; see for instance \cite{ponsotschomerusteschner, stanciu,
ooguriparktannenhauser, Griegos, parnachevsahakyan, ilnuevo3, ribault2,
ribault3} and references therein.

The case $m=b^{2}$ also yields the $H_{+}^{3}$ WZNW\ theory with level $%
k=b^{+2}+2$. The fact that the WZNW theory is doubly represented in the
family \{$\mathcal{T}_{mb}$\} is associated to Langlands duality \cite%
{giribetnakayamanicolas}. In the free field approach this is related to the
existence of a "second" screening operator. In this context, free field
calculations using the Wakimoto representation with $m>1$ were already
discussed in \cite{GN3}. In fact, the computation we performed this paper
can be regarded as a generalization of the one in \cite{GN3} to the geometry
of the disk and to the case where $m$ is not necessarily equal to $b^{+2}+2$.
Furthermore, one could feel tempted to go a step further and conjecture that
a relation similar to (\ref{torbe})-(\ref{asdefined}) also holds for one-point functions
in the theory ${\mathcal T}_{b,m}$ with arbitrary (not necessarily integer) value of $m$.


Other case of interest is the theory for $m=2$. In this case, correlation
functions on the sphere were shown to obey third order differential
equations that are associated to existence of singular vectors in the modulo 
\cite{ribault1}. This raises the question as to whether the existence of
singular vectors could be used to compute observables in the theory with
boundaries by means of the bootstrap approach or some variation of it. It
could be also interesting to attempt to compute observable in the presence
of a boundary by using a free field representation similar to that proposed
in \cite{Yo} to describe the $m=1$ theory. Such free field representation
amounts to describe the $H_{+}^{3}$ WZNW theory as a $c<1$ perturbed
conformal field theory coupled to Liouville theory, resorting to the $%
H_{+}^{3}$ WZNW-Liouville correspondence (see also \cite{hikidaschomerus3}).

Now, let us analyze here some properties of the one-point function for $m\in 
\mathbb{Z}$ we have computed in (\ref{93}). Important information is
obtained from studying how $\Omega _{j}^{(m,b)}$ transforms under certain
changes in the set of quantum numbers $(m,b,j)$ that leave the conformal
dimension $h_{j}$ unchanged. Looking at this gives important information
about the symmetries of the theory. But, first, a few words on the scaling
properties of the one-point function (\ref{93}): The reason why we
previously said that the precise value of the overall factor $\Omega _{0}$
in (\ref{L1}) and (\ref{nueve}) was "irrelevant" was that, by shifting the
zero-mode of the field $\phi $ one easily introduces a KPZ scaling $\lambda $
in the bulk expectation value, and this yields%
\begin{equation}
\Omega _{j}^{(m,b)}\rightarrow \lambda ^{(2j+1+(m-1)b^{-2})/2}\text{ }\Omega
_{j}^{(m,b)}.
\end{equation}%
Then, since $\Omega _{0}$ also goes as a power $n/2=-(2j+1+(m-1)b^{-2})/2$,
its value can be absorbed and conventionally fixed to any (positive value).
In turn, we prefer to write (\ref{93}) by replacing 
\begin{equation}
\Omega _{0}\left( \pi \frac{\Gamma (1-b^{2})}{\Gamma (1+b^{2})}\right)
^{(2j+1+(m-1)b^{-2})/2}\rightarrow \left( \lambda \pi \frac{\Gamma (1-b^{2})%
}{\Gamma (1+b^{2})}\right) ^{(2j+1+(m-1)b^{-2})/2}  \label{99}
\end{equation}

Now, we are ready to study the reflection properties of (\ref{93}). Using
properties of the $\Gamma $-function, it is easy to verify that the
following relation holds%
\begin{equation}
\Omega _{j}^{(m,b)}\Omega _{-1-j-\frac{m-1}{b^{2}}}^{(m,b)}=R_{j}^{(m,b)},%
\quad \text{with}\quad R_{j}^{(m,b)}=\frac{1}{b^{2}}\left( \lambda \pi \frac{%
\Gamma (1-b^{2})}{\Gamma (1+b^{2})}\right) ^{2j+1+\frac{m-1}{b^{2}}}\frac{%
\gamma (2j+1+(m-1)b^{-2})}{\gamma (-(2j+1)b^{2}-(m-1))},\text{ }
\label{torbe2}
\end{equation}%
where $\gamma (x)=\Gamma (x)/\Gamma (1-x)$. It is remarkable that the
reflection coefficient $R_{j}^{(m,b)}$, which is given by the two-point
function on the sphere, arises in this expression. This generalizes what
happens in Liouville field theory and in the $H_{3}^{+}$ WZNW model, and
this is related to the fact that one eventually associates fields $\Phi _{j}$
and fields $R_{j}^{(m,b)}\Phi _{-1-j-b^{-2}(m-1)}$.

Other functional property of the one-point function that is interesting to
analyze is how it behaves under duality transformation \thinspace $%
b\rightarrow 1/b$. Actually, one can show that (\ref{93}) obeys%
\begin{equation}
\Omega _{j}^{(m,b)}=\Omega _{b^{2}(j+1-b^{-2})}^{(mb^{-2},b^{-1})},
\end{equation}%
provided the KPZ scaling parameter $\lambda ^{\prime }$ associated to the
function on the right hand side relates to that of the function on the left
hand side through%
\begin{equation}
\left( \lambda ^{\prime }\pi \frac{\Gamma (1-b^{-2})}{\Gamma (1+b^{-2})}%
\right) ^{b^{2}}=\lambda \pi \frac{\Gamma (1-b^{2})}{\Gamma (1+b^{2})}.
\end{equation}

Last, let us mention another interesting problem that involves the theories $%
\mathcal{T}_{m,b}$ formulated on closed Riemann surfaces and that is in some
sense related to the path integral techniques we discussed here. This is the
problem of trying to use the path integral approach developed in \cite%
{schomerus} to define higher genus correlation functions for $m\in \mathbb{Z}
$, relating higher genus correlators in $\mathcal{T}_{m,b}$ to higher genus
correlators in Liouville theory. Correlation functions on closed genus-$g$ $%
n $-punctured Riemann surfaces in the theories $\mathcal{T}_{m,b}$ could be
relevant to describe higher $m$-monodromy operators in N=2 four-dimensional
superconformal field theories, according to the recently proposed
Alday-Gaiotto-Tachikawa conjecture \cite{AGT1, AGT2, AGT3}; this is because $%
n$-point functions in $\mathcal{T}_{m,b}$ are in correspondence with $%
(2n+2g-2)$-point functions in Liouville theory including $2g-2+n$ degenerate
fields $V_{a=-\frac{m}{2b}}$. Studying the relevance of the theories defined
in \cite{ribault1} for the AGT construction is matter of future
investigation.

\begin{equation*}
\end{equation*}

This work was supported by UBA, ANPCyT, and CONICET through grants UBACyT
X861, UBACyT X432, PICT-2007-00849 and PIP-2010-0396. The authors thank
Lorena Nicol\'{a}s and Sylvain Ribault for discussions. They also thank the
referee of JHEP for thoughtful remarks on the manuscript; in particular, for
suggesting the definition of the boundary parameter $r$ as in
(\ref{asdefined}). G.G. thanks Yu Nakayama for 
previous collaborations in the
subject.

\end{document}